# High-resolution detection of neutral oxygen and non-LTE effects in the atmosphere of KELT-9b

Francesco Borsa[1,*], Luca Fossati[2], Tommi Koskinen[3], Mitchell E. Young[4], and Denis Shulyak[5]

[1]INAF - Osservatorio Astronomico di Brera, Via E. Bianchi 46, 23807 Merate (LC), Italy
[2]Space Research Institute, Austrian Academy of Sciences, Schmiedlstrasse 6, A-8042 Graz, Austria
[3]Lunar and Planetary Laboratory, University of Arizona, 1629 East University Boulevard, Tucson, AZ 85721-0092,USA
[4]Department of Physics, University of Oxford, Denys Wilkinson Building, Keble Road, Oxford, OX1 3RH United Kingdom
[5]Instituto de Astrofísica de Andalucía - CSIC, c/ Glorieta de la Astronomía s/n, 18008 Granada, Spain
[*]francesco.borsa@inaf.it

**Oxygen is a constituent of many of the most abundant molecules detected in exoplanetary atmospheres and a key ingredient for tracking how and where a planet formed[1]. In particular, the OI 777.4 nm triplet is used to probe airglow and aurora on the Earth[2] and the oxygen abundance in stellar atmospheres[3,4,5,6], but has not been detected in an exoplanet atmosphere before. We present a definite ground-based detection of the neutral oxygen 777.4 nm triplet lines in the transmission spectrum of the ultra-hot Jupiter KELT-9b[7], the hottest known giant planet. The synthetic spectrum computed employing novel non-local thermodynamic equilibrium (NLTE) radiative transfer calculations[8] matches the data significantly better compared to the one computed assuming local thermodynamic equilibrium. These NLTE radiative transfer calculations imply a mass-loss rate of $10^8$-$10^9$ kg s$^{-1}$, which exceeds the lower limit of $10^7$-$10^8$ kg s$^{-1}$ required to facilitate the escape of oxygen and iron from the atmosphere. Assuming a solar oxygen abundance, the NLTE model points towards the need of microturbulence and macroturbulence broadening of 3.0±0.7 km s$^{-1}$ and 13±5 km s$^{-1}$, respectively, indicative of the presence of fast winds in the middle and upper atmosphere. Present and upcoming high-resolution spectrographs will allow the detection in other exoplanets of the 777.4 nm OI triplet, which is a powerful tool to constrain the key characteristics of exoplanetary atmospheres when coupled with forward modelling accounting for NLTE effects.**

The atmosphere of KELT-9b[9,10,11,12,13,14,15,16] is characterised by a temperature inversion[17,15,18] with an upper atmospheric temperature of the order of 8000-10000 K[19,11,12,18,8]. In particular, Ref.[8] computed self-consistently the temperature-pressure (TP) profile and composition of the atmosphere of KELT-9b accounting for non-local thermodynamic equilibrium (NLTE) effects, which lead to deviations of the excitation and ionisation states of atoms and molecules from those predicted by the Boltzmann and Saha equations, respectively. They predicted an upper atmospheric temperature about 2000 K hotter than the one computed assuming local thermodynamic equilibrium (LTE). The NLTE transmission spectrum based on this TP profile provides an excellent match to the observed hydrogen Balmer lines[8]. Furthermore, it showed that, as a result of the high atmospheric temperature and of NLTE effects, the triplet of neutral oxygen at ~777.4 nm should be almost as strong as the previously detected Hα Balmer line. We analysed public data downloaded from the Calar Alto archive, that were taken with the CARMENES spectrograph during three transits of KELT-9b. CARMENES[20] is a fiber-fed high-resolution echelle spectrograph installed on the 3.5 m telescope at Calar Alto observatory, whose optical arm covers the 520-960 nm wavelength range with an average resolving power of about 94,600, and is thus well-suited for detecting neutral oxygen at the position of the 777.4 nm triplet.

By using a differential technique and comparing observations taken in- and out-of-transit, we can extract the transmission spectrum of the planetary atmosphere. For each of the three available transit datasets, we independently extracted the planetary transmission spectrum. A telluric correction was not required in the wavelength range covered by the OI triplet (see Methods). We created a stellar spectrum reference by averaging the out-of-transit measurements after moving to the stellar rest frame, then all the spectra were normalised for this reference. All the full-in-transit spectra (i.e., when the planetary disk is completely contained inside the stellar disk) of the three transits were then moved to the planetary rest frame and averaged to create the final planetary transmission spectrum (Fig. 1).

Spurious stellar artifacts, such as Rossiter-McLaughlin (RM) effect or center-to-limb variations can contaminate the transmission spectrum, causing line profile deformation[21] and false detections[22]. We verified that this is not the case here by taking these effects into account in our analysis (Methods). We further checked that the oxygen absorption signal originates from the planet by verifying that the absorption signal follows the expected velocity of the planet. This was done by using the 2D tomographic technique after averaging the OI triplet in velocity space (Fig. 2a) and by computing the $K_p$-$V_{sys}$ map, leading to a detection at a signal-to-noise ratio of 9.2 compatible with the planetary rest frame (Fig. 2b and Methods).

We fitted the line profiles present in the transmission spectrum with Gaussian functions in a Bayesian framework (Methods). We derived for the oxygen triplet an average line contrast of 0.255±0.015 %, and a full width at half maximum (FWHM) of 20.7±1.8 km s$^{-1}$. The line contrast corresponds to an average effective planetary radius of ~1.17 $R_P$. For a tidally locked rotating atmosphere, such as that expected for KELT-9b, an absorption feature at ~1.17 $R_P$ should have a FWHM of ~13 km s$^{-1}$. The larger FWHM obtained from the Gaussian fit points to the presence of additional broadening mechanism(s).

We compared the observations with the synthetic transmission spectrum computed with our NLTE model of the atmosphere, assuming a solar oxygen abundance[8]. Here, our model accounts for the spectral resolution of the instrument and planetary rotation (assuming tidal locking). In line with the data, the model shows prominent absorption by the triplet, but the observed line profiles are broader than the model lines (Methods). To find a better match between the observation and the model, we computed transmission spectra in the same way as described in Ref.[8] applying different microturbulence ($v_{mic}$; i.e., velocity of gas on a scale smaller than the pressure scale height) and macroturbulence ($v_{mac}$; i.e., velocity of gas on a scale larger than the pressure scale height) velocity values. We employed a $X^2$ minimisation routine (Methods) to identify the best fitting $v_{mic}$ and $v_{mac}$ values and obtained $v_{mic}$=3.0±0.7 km s$^{-1}$ and $v_{mac}$=13±5 km s$^{-1}$ (Fig. 3). The constraint on the microturbulence velocity is tighter than that on the macroturbulence velocity, because $v_{mic}$ variations also affect the strength of the lines. Furthermore, within 2$\sigma$, the observations are also consistent with zero macroturbulence velocity. We note that assuming a higher abundance of oxygen is an alternative to increasing the microturbulence velocity, and thus a somewhat higher oxygen abundance in the model, and/or a lower ionization fraction of oxygen, could also help to improve the fit. A detailed exploration of the parameter space, however, is beyond the scope of the present work. Also, the value of $v_{mic}$ that we obtained agrees well with that used by Ref.[11] to fit the CaII infrared triplet in the planetary transmission spectrum assuming solar abundance.

We also compared the observations with a model obtained assuming LTE in computing both atmospheric TP profile and transmission spectrum (Fig. 1, from Ref.[8]), further broadening it accounting for $v_{mic}$ and $v_{mac}$ values of 10±2 km s$^{-1}$ and 0±7 km s$^{-1}$, respectively, that we estimated as done for the NLTE spectrum. The LTE profile is significantly weaker and a worse fit compared to the spectrum computed in NLTE, as shown by the respective $X^2$ values of 275 (NLTE) and 297 (LTE), with a preference for the NLTE model at the 4.3σ level (see Methods). Unfortunately, the current modelling setup of Ref.[8] does not enable fine tuning of the abundance of single elements, but the fact that the NLTE transmission spectrum fits well the hydrogen Balmer lines (Methods) and the OI triplet is strong indication that the model is capable of reproducing the average atmospheric properties in the region probed by these lines with solar abundances and that the OI triplet can be used to study NLTE effects in the atmospheres of ultra-hot Jupiters.

Understanding the possible origin of the macroturbulence velocity requires a deeper look at atmospheric structure and circulation. Figure 4 shows the sound speed and Jeans escape parameter as a function of atmospheric structure (Methods). Here, the effective Jeans escape parameter is based on the gravitational potential difference between the given pressure level and the Roche lobe boundary (e.g., Ref.[23]). Radial expansion at a velocity of 5-6 km s$^{-1}$ due to atmospheric escape in the line formation region (5 x 10$^{-9}$ - 5 x 10$^{-5}$ bar) could explain the observed broadening, but is unlikely. The effective Jeans escape parameter reduces to zero at the Roche lobe at a pressure of about 10$^{-10}$ bar. To a good approximation, the atmosphere escapes through the Roche lobe around the L1 and L2 points at sound speed. Because gas escaping at other latitudes around the planet is also directed towards the L1 and L2 points, we can estimate the global mass loss rate to within an order of magnitude by assuming that escape proceeds at sound speed through the entire Roche lobe surface[24,25]. Based on this assumption, our model TP profile implies a mass loss rate of 10$^8$-10$^9$ kg s$^{-1}$, which translates to a radial velocity of the order of 100 m s$^{-1}$ in the line formation region. This exceeds the mass loss rate that would be required to enable the escape of neutral oxygen and iron from the atmosphere (Methods), but it is not sufficient to explain the broadening of the observed lines. A radial velocity of the order of 5 km s$^{-1}$ in the line formation region would imply a much higher mass loss rate, with potentially disastrous consequences for the planet. Global circulation provides a more attractive alternative. Current circulation models for ultra-hot Jupiters predict wind speeds of several km s$^{-1}$, even for planets that are substantially cooler than KELT-9b[26]. Such wind speeds are in good agreement with the observations.

The OI detection in the atmosphere of KELT-9b shows that atomic oxygen can be detected and measured from the ground at optical wavelengths and not exclusively from space in the far-ultraviolet[27,28]. Furthermore, the OI triplet at ~777.4 nm lies in a spectral window almost unaffected by telluric absorption and is accessible by several ground-based instruments attached to large telescopes, possibly enabling to detect oxygen from the ground also for other exoplanets orbiting stars fainter than KELT-9. Accounting for the relevant physical processes, including NLTE effects, enables the comparison of forward models with the observations to constrain the key characteristics of exoplanetary atmospheres, including the abundance of oxygen, mass loss, and velocity state, similarly to what is routinely done with stellar atmospheres.

**Acknowledgements** F.B. acknowledges support from PLATO ASI-INAF agreement n. 2015-019-R.1-2018. This research has made use of the Spanish Virtual Observatory (http://svo.cab.inta-csic.es) supported by the MINECO/FEDER through grant AyA2017-84089.7. T.K. acknowledges support by the NASA Exoplanet Research Program grant 80NSSC18K0569. D.S. acknowledges the financial support from the State Agency for Research of the Spanish MCIU through the "Center of Excellence Severo Ochoa" award to the Instituto de Astrofísica de Andalucía (SEV-2017-0709). M.E.Y. acknowledges funding from the European Research Council (ERC) under the European Union's Horizon 2020 research and innovation program under grant agreement No 805445.

**Code availability**
The spectral reduction was done with a self-written IDL script. The stellar model spectrum used for the stellar contamination impact was obtained with the Spectroscopy Made Easy tool, which is publicly available from http://www.stsci.edu/~valenti/sme.html. The DE-MCMC routines were taken from EXOFAST, publicly available at https://github.com/jdeast/EXOFASTv2.

**Data availability** Data used in this work are publicly available from the Calar Alto archive at http://caha.sdc.cab.inta-csic.es/calto/.

**Author contributions statement** F.B. carried out the data analysis. L.F. and T.K. performed the theoretical calculations presented in this work. F.B., L.F., and T.K. contributed to the writing of the manuscript. All authors contributed to the interpretation of the data and the results.

**Competing interests** The authors declare that they have no competing interests.

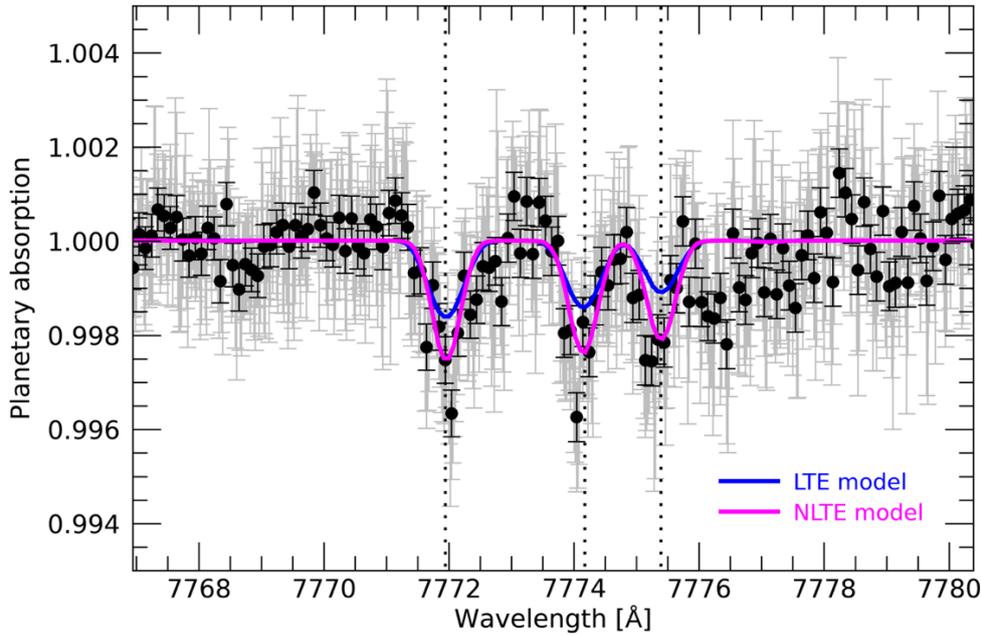

**Figure 1.** Transmission spectrum of KELT-9b around the OI triplet. The magenta line shows the NLTE model of Ref.[8] broadened with the best fit microturbulence ($v_{mic}$) and macroturbulence ($v_{mac}$) velocity, while the blue line shows the same for the LTE model (also from Ref.[8]). Grey lines show individual measurements with related uncertainties, black points show an 0.1 Angstrom binning. The vertical dotted lines mark the expected position of the OI triplet lines.

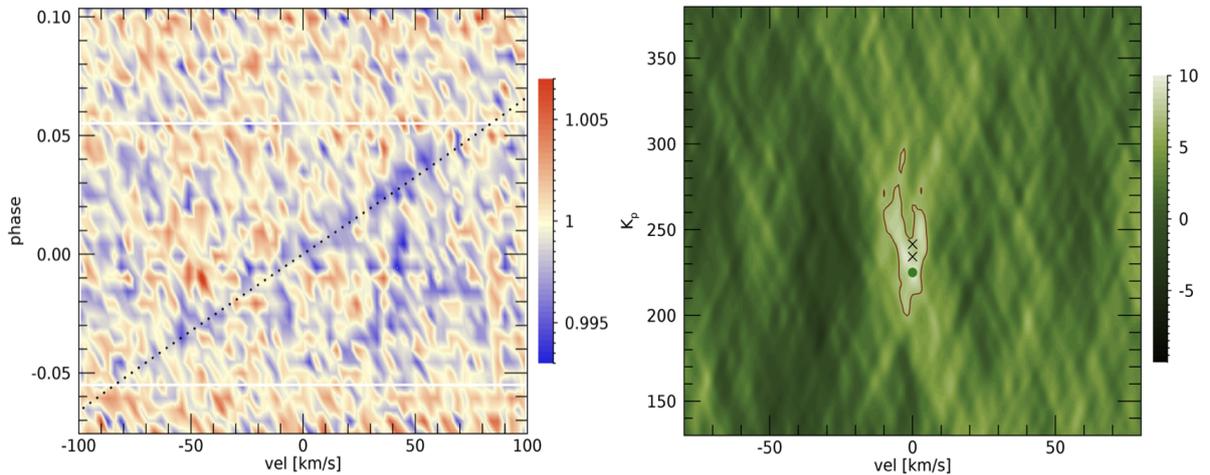

**Figure 2. a,** 2D contour map of the oxygen triplet averaged in velocity space, in the stellar rest frame. The black dotted line shows the expected motion of the planetary signal. The white horizontal lines mark the beginning and end of the transit. The colour scale shows the relative flux variation. **b,** $K_p$-$V_{sys}$ map for the OI triplet. Crosses mark the position of the $K_p$ (the RV semi-amplitude of the planetary Keplerian motion) determined by Refs.[14,15]. The green dot marks the best $K_p$ position in the map. The solid line shows the $3\sigma$ confidence interval. The colour scale shows the signal-to-noise ratio of the detection.

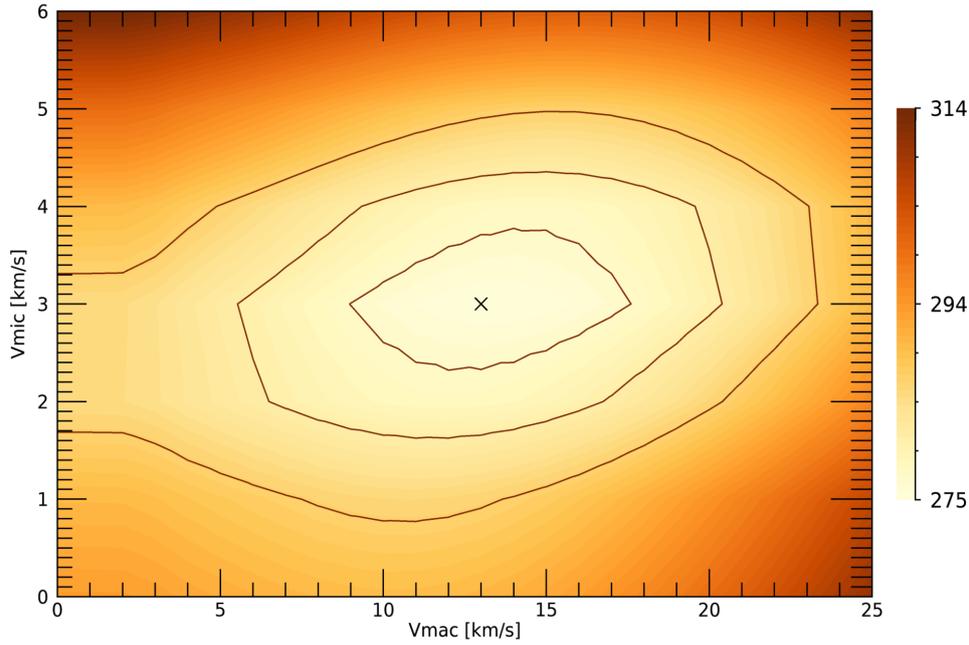

**Figure 3.** $X^2$ minimization map of the NLTE synthetic spectrum with contour lines of $1\sigma$, $2\sigma$ and $3\sigma$ confidence intervals for the values of the microturbulence ($v_{mic}$; $3.0\pm0.7$ km s$^{-1}$) and macroturbulence ($v_{mac}$; $13\pm5$ km s$^{-1}$) velocities.

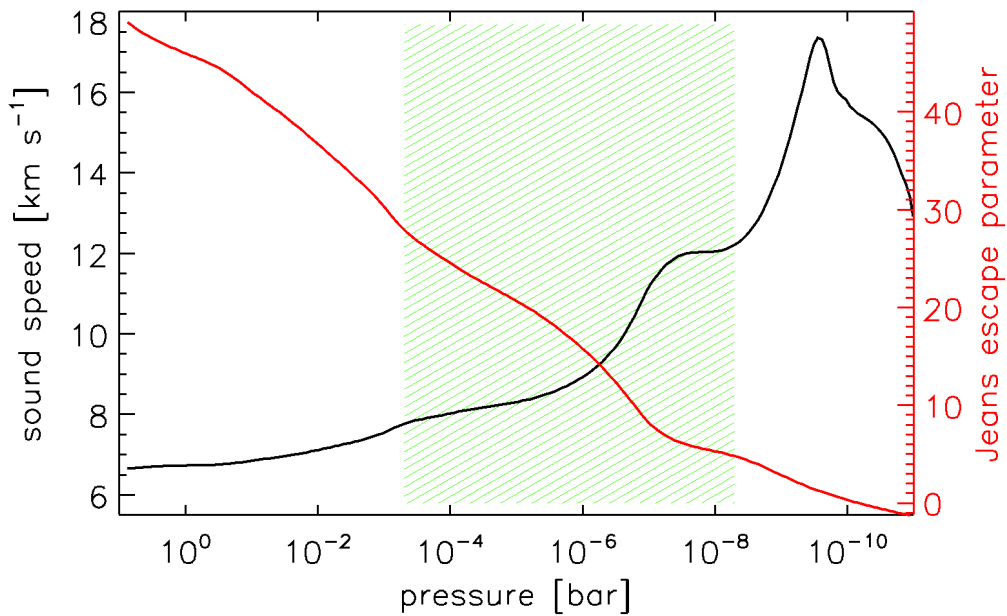

**Figure 4.** Sound speed (black) and Jeans escape parameter (red) as a function of pressure extracted from the NLTE atmospheric structure profile presented by Ref.[8]. The shaded area indicates the main formation region of spectral lines lying in the optical.

## Methods

### Data sample and reduction

We analysed spectroscopic public data downloaded from the Calar Alto archive (http://caha.sdc.cab.inta-csic.es/calto/), which contains data covering three transits of KELT-9b taken in the nights of 2017-08-06, 2017-09-21, and 2018-06-16 with CARMENES[20]. This is a fiber-fed high-resolution echelle spectrograph installed on the 3.5 m telescope at Calar Alto observatory. Its optical arm covers the ~520-960 nm wavelength range with an average resolving power of about 94,600. The data of each night (57, 56 and 140 spectra, respectively) cover the transit of the planet. The exposure times used were different, varying from 111 sec to 400 sec. The signal-to-noise ratio (S/N) in the order 39 of the spectrograph, where the oxygen triplet lies, ranges from 30 to 120.

We analysed the reduced archival data, that were extracted with the instrument pipeline version 2.01[29]. The pipeline, after correcting for bias, flat field, and cosmic rays, performs a flat-relative optimal extraction[30] and wavelength calibration[31]. The output is produced in order-by-order maps, and consists of flux, flux-uncertainties, and wavelength (in vacuum and Earth rest frame). Before performing our analysis, for each night we discarded spectra taken at an airmass higher than 2 and with a S/N lower than half of the average S/N of the night (4, 15 and 4 spectra, respectively).

### Transmission spectrum extraction

We focused just on the order number 39 of the spectrograph, after shifting it from vacuum to air wavelengths, and in particular around the 775-780 nm wavelength range, which contains the oxygen triplet (777.194, 777.417, 777.539 nm, in air wavelengths). Because of the fast stellar rotation (Vsini~110 km s$^{-1}$), thus line blending, the stellar OI triplet appears as one broad feature (Supplementary Fig. 1). Throughout, we considered the system parameters given by Ref.[16].

We extracted the transmission spectrum by following independently for each night a procedure based on that presented by Ref.[32]. The spectra are first shifted to the stellar rest frame by using the barycentric Earth radial velocity given by the pipeline and a Keplerian model of the system, then each spectrum is continuum normalised by performing a linear fit between two wavelength ranges adjacent to the region of interest (776.0-776.3 nm and 778.2-778.4 nm). Finally, we divided all spectra by the average stellar spectrum created combining the out-of-transit spectra. At the end, we moved into the planetary rest frame by shifting all the residual spectra for the theoretical planetary radial velocity, using the planetary orbital $K_P$ calculated from the orbital motion (247.28 km s$^{-1}$, Ref.[16]). The final rebinning is done on a wavelength step of 0.02 Angstrom, chosen for being close to the mean resolution step in the range of interest. We then created the transmission spectrum with a weighted average of all full-in-transit residual spectra.

### Telluric correction

We performed telluric correction using a scaling relation between airmass and telluric line strength[33,34,35], rescaling all the normalised stellar spectra as if they were obtained at the airmass of the center of the transit. We remark that no telluric line was standing beyond the noise level in the considered wavelength range. We further checked for telluric lines in this wavelength range in archival CARMENES spectra of the telluric standard 109 Vir, without finding any. We thus decided not to apply our telluric correction to avoid increasing the noise in the final transmission spectrum, but the final results are independent from this choice. We further checked if any water micro-telluric could be present in the wavelength range we analysed employing the ESO Sky Model Calculator (https://www.eso.org/observing/etc/bin/gen/form?INS.MODE=swspectr+INS.NAME=SKYCALC). Rapid water vapour changes in the Earth atmosphere could make the depth of the telluric lines not follow the correlation with airmass, and could thus be missed by our telluric correction. We created models by selecting two very different values of precipitable water vapour (5 and 20 mm), and simulated variability between these values during the transits, shifting the telluric models using the values of $V_{sys}$ and barycentric Earth radial velocity for the different nights and convolving them with the instrumental broadening. Water micro-tellurics could partly overlap with the reddest line of the OI triplet, we can thus not completely exclude that this line could be partly contaminated. We however confirm that the first two lines of the triplet are instead free also from water micro-tellurics (Supplementary Fig. 2).

Given the position of these water micro-tellurics in the transmission spectrum (about 777.6-777.7 nm), we can argue that their possible variation could be the cause of the imperfect normalisation on the right hand side of the triplet. As a test to see the possible impact on our analysis, we recomputed the $X^2$ map of Fig. 3 by comparing the planetary atmospheric models and the observed transmission spectrum only blueward of 7775 Angstrom, thus excluding the third line of the OI triplet, obtaining that our final results remain unchanged, confirming the $v_{mic}$ and $v_{mac}$ values we found.

As for telluric emission, we checked fiber B of the instrument, which was pointing at the sky during the observations. No emission is present beyond the noise when co-adding the spectra in the telluric reference frame for each night.

**Stellar contamination in the transmission spectrum**

The host star is not an homogeneous disk, but has a surface brightness which rotates and changes as a function of the distance from the center. Effects such as center-to-limb variation (CLV) and stellar rotation (the RM effect) may lead to spurious signals in the transmission spectrum, possibly causing false detections[22] and wrong line-profile estimates[21].

We modeled these effects using the approach of Ref.[36] in which the star is modeled as a disk divided in sections of 0.01 $R_S$, each section having its value of projected rotational velocity (propagated using the Vsini value) and μ (μ=cos(θ), with θ being the angle between the normal to the stellar surface and the considered line of sight). A spectrum is assigned to each point of the grid, by quadratically interpolating on μ and Doppler-shifting the model spectra created using Spectroscopy Made Easy[37] with ATLAS12 stellar atmospheric models (assuming solar abundances and LTE) and the line list from the VALD database[38]. The model spectra are created for 21 different μ values and with null rotational velocity, and adapted to the resolving power of the instrument.

The stellar spectrum for each orbital phase during the transit of the planet is then calculated as the average of the non-occulted modeled sections. We then modeled the CLV and RM effect influence on the transmission spectrum by dividing all spectra by a master out-of-transit stellar spectrum, moving into the planetary rest frame and averaging all the full-in-transit spectra.

The magnitude of the stellar contamination given by CLV and RM effect is found to be well within the noise of the observations, demonstrating that the oxygen detection is not an artifact (Supplementary Fig. 3). Despite this, we applied the correction, normalising the transmission spectrum for the stellar contamination model.

## 2D tomographic confirmation

Spurious stellar contamination beyond the one given by CLV and rotation may still exist, as the stellar line profiles and/or not perfectly corrected tellurics may show variability during transit and cause spurious effects[39,21,40,41]. We thus verified that the signal belongs to the planetary rest frame by looking at the 2D tomographic map, where it is clear that the absorption signal follows the planetary rest frame and not the stellar one. To increase the S/N, we passed from wavelengths to velocities by using Doppler's law, centered on each line of the oxygen triplet, and then averaged the three lines. Although aliases are present due to the proximity of the lines, the detected absorption is of planetary nature (Fig. 2a).

We further checked the rest frame of the signal by computing the $K_p$-$V_{sys}$ map (Fig. 2b). This is created by cross-correlating the residual spectra in the stellar rest frame (i.e. the spectra after dividing by the average out-of-transit stellar spectrum) with a template modelled with three normalised Gaussian functions, centered to the rest frame of the OI triplet and considering the FWHM given by the instrumental resolution. We then averaged the in-transit cross-correlation functions after shifting them in the planetary rest frame, for a range of $K_p$ values from 100 to 400 km s$^{-1}$ in steps of 1 km s$^{-1}$. This is done by subtracting the planetary radial velocity calculated for each spectrum as $V_p = K_p \times \sin 2\pi\varphi$, with $\varphi$ the orbital phase. We thus created the $K_p$ versus $V_{sys}$ map to verify the real planetary origin of the signal. We evaluated the noise by calculating the standard deviation of the map taken in the range $K_p$=[150:360] km s$^{-1}$ and $V_{sys}$=[20:75] km s$^{-1}$, far from where any stellar or planetary signal is expected. We note that this noise sampling is not perfect, as the map shows strong aliasing and correlated noise. While the $K_P$ can not be determined with high precision since the model only has three lines, the best $K_P$ position of the signal is compatible with the one determined in other works analysing metals in the atmosphere of KELT-9b[14,15], thus further validating the planetary nature of the OI signal. The detection is at S/N=9.2.

We also performed a comparison between the in-trail (|v|<6 km s$^{-1}$) and out-of-trail (|v|>20 km s$^{-1}$) values shown in Fig. 2 in the planetary rest frame. A Welch t-test on the data rejects the hypothesis that the two samples are drawn from the same parent distribution at the 8.4σ level, although we remind the reader that the strong aliasing slightly biases this statistics making the distributions not perfectly Gaussian.

## Planetary absorption lines Gaussian fit

The final transmission spectrum presents an imperfect normalisation just on the right of the oxygen triplet, i.e. at ~777.65 nm, that is possibly caused by micro-telluric contamination (Sect. Telluric correction). Indeed, the synthetic spectrum of Ref.[8], which accounts for the lines of all metals up to Zn, does not show any feature at this wavelength.

We fitted the sum of three Gaussian functions to the OI absorption features (Supplementary Fig. 4), imposing the same full width at half maximum (FWHM) for each line to avoid the imperfect normalisation redward of the triplet from influencing the fit of the 777.539 nm line. The fit was performed in a Bayesian framework employing a differential evolution Markov chain Monte Carlo (DE-MCMC) technique[42,43], running ten DE-MCMC chains of 100,000 steps and discarding the burn-in. The medians and the 15.86% and 84.14% quantiles of the posterior distributions were taken as the best values and 1σ uncertainties. The results of the fit are shown in Supplementary Table 1. The line contrasts were translated into an average effective planetary radius of $R_{eff}$ ~1.17 $R_P$ by assuming $(R_{eff}/R_P)^2 = (\delta+h)/\delta$, where $\delta$ is the transit depth and $h$ is the line contrast.

Unfortunately we are not sensitive to any possible asymmetry of the lines given the S/N available. When fitting the lines with skewed Gaussian functions, the skew parameter remains unconstrained.

**Line broadening of the model**

We compared the observed transmission spectrum of the OI triplet to the NLTE synthetic transmission spectrum that guided our observations[8]. After having added instrumental (using a Gaussian kernel with FWHM of 3.17 km s$^{-1}$) and rotational broadening (assuming tidal locking at a radius of 1.17 R$_P$ and a rotational profile calculated as in Ref.[39]) to the synthetic line profiles, the synthetic spectrum appeared to be narrower than the observation. Line broadening due to the exposure time, caused by the fact that the planet moves during each exposure, has a negligible impact on the overall broadening (i.e. ~4.7, 3.5, and 1.3 km s$^{-1}$ on average for exposure times of 400, 300, and 111 sec, respectively).

Therefore, we recomputed further transmission spectra models in the same way as described in Ref.[8] applying different microturbulence velocity ($v_{mic}$) values ranging between 1 and 14 km s$^{-1}$ (in steps of 1 km s$^{-1}$), further adding macroturbulence velocity ($v_{mac}$) broadening with $v_{mac}$<25 km s$^{-1}$ (in steps of 1 km s$^{-1}$), finally looking for the $v_{mic}$ and $v_{mac}$ pair leading to match the observations by using X$^2$ minimization. We obtained best fitting values of $v_{mic}$= 3.0 ± 0.7 km s$^{-1}$ and $v_{mac}$=13 ± 5 km s$^{-1}$ (Figure 3). Remarkably, the inclusion of these broadening values does not worsen the fit of the hydrogen Balmer lines (Supplementary Fig. 5).

We followed the same procedure on the LTE transmission spectrum of Ref.[8] to estimate the $v_{mic}$ and $v_{mac}$ values best fitting the observations obtaining 10±2 km s$^{-1}$ and 0±7 km s$^{-1}$, respectively (Supplementary Fig. 6). However, the LTE transmission spectrum minimising the X$^2$ value is a significantly worse fit to the observation compared to the NLTE synthetic spectrum. We compared the obtained X$^2$ values of 275 (NLTE model) and 297 (LTE model) by using a likelihood ratio test. With two-degrees of freedom ($v_{mic}$ and $v_{mac}$), we find a p-value of 1.67x10$^{-5}$, that corresponds to a 4.3$\sigma$ preference for the NLTE model when assuming that the uncertainties have been correctly estimated.

**Mass loss rate**

To aid the interpretation of these results, we employed the atmospheric model presented by Ref.[8] to estimate the mass loss rate. Figure 4 shows the effective Jeans escape parameter and sound speed in the atmosphere based on our model TP profile. We remind the reader that the sound speed represents the largest possible atmospheric radial velocity below the L1 point. The effective Jeans escape parameter is

$$X = \frac{m\Delta\varphi}{kT}$$

where Δφ is the gravitational potential difference between the pressure level and the Roche lobe. At the Roche lobe, X = 0 and we can estimate the mass loss rate as

$$\frac{dM}{dt} \simeq 4\pi YRL1\rho \frac{C}{2\sqrt{\pi}}$$

where R$_{L1}$ is the radius at the L1 point, Y = 2R$_{L1}$/3 is the polar radius and C is the thermal speed. The L1 point in our model is located at 2.6 R$_p$ with a pressure of 1.17 x 10$^{-10}$ bar where the thermal speed is 14 km s$^{-1}$ and mass density is 1.17 x 10$^{-13}$ kg m$^{-3}$. The resulting value for the mass loss rate is 5 x 10$^8$ kg s$^{-1}$. We expect this estimate to be accurate roughly to an order of magnitude[24,25].

The cross-over mass concept[44] can be used to derive the minimum mass loss rate of hydrogen required to enable atomic oxygen and iron to escape the atmosphere[45]:

$$\frac{dM}{dt} \simeq 4\pi m_H^2 G M_p (M_c - 1) \frac{nD}{kT}$$

where $m_H$ is the mass of hydrogen, $M_c$ is the mass of the heavy element in units of $m_H$ and $nD$ is the product of the total number density and the mutual diffusion coefficient. Using values of $nD \simeq 1.7 \times 10^{22}$ m$^{-1}$ s$^{-1}$ for O-H$^+$ collisions or $nD \simeq 1.4 \times 10^{22}$ m$^{-1}$ s$^{-1}$ for O-H collisions gives a limiting mass loss rate of 2-3 x 10$^7$ kg s$^{-1}$ for neutral oxygen. The value obtained for iron is about 10$^8$ kg s$^{-1}$.

We note that our atmosphere model is hydrostatic and does not include escape. The atmosphere begins to deviate from hydrostatic equilibrium and adiabatic cooling due to expansion becomes significant once the outflow velocity reaches a substantial fraction of the sound speed. Based on a mass loss rate of 5 x 10$^8$ kg s$^{-1}$ and atmospheric structure predicted by our model, this would happen at 10$^{-10}$-10$^{-9}$ bar, given that $4\pi\rho v r^2$ is constant with radius. This location is close to our upper boundary, and above the region probed by the oxygen line profile. Detailed models of escape that include heavy elements are required to calculate the mass loss rate more precisely and interpret transit signatures that probe higher altitudes in the upper atmosphere.

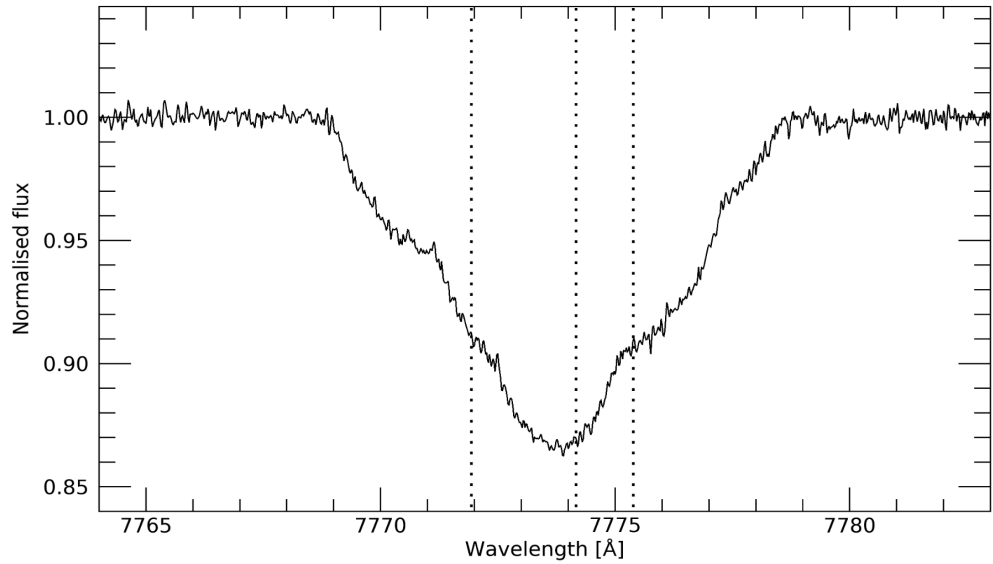

**Supplementary Figure 1.** Normalised master stellar spectrum in the region of the oxygen triplet. The vertical dotted lines mark the position of the three lines of the triplet.

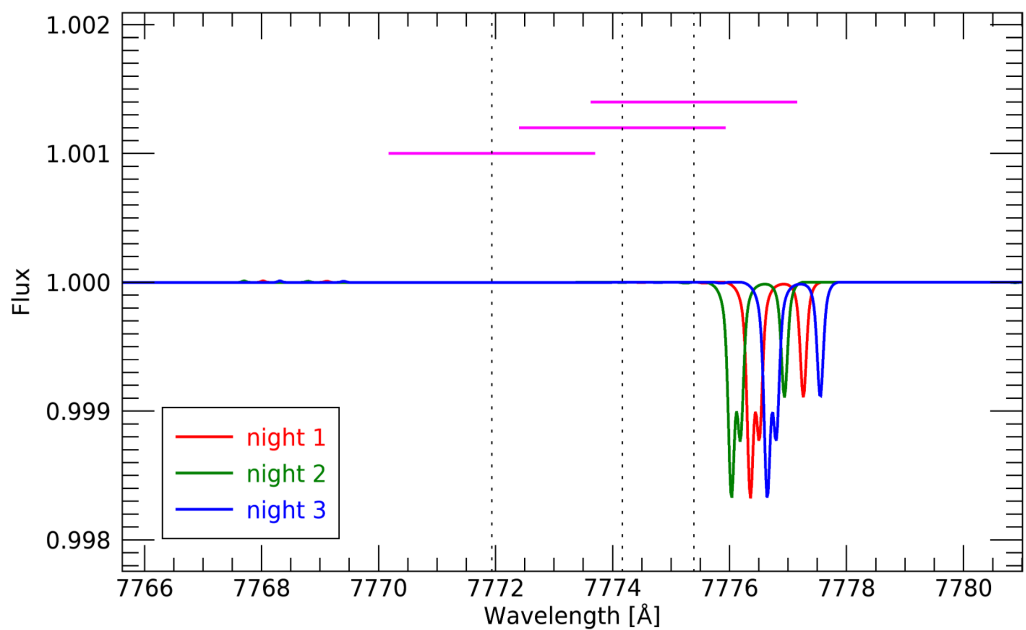

**Supplementary Figure 2.** Position of telluric contamination possibly caused by strong precipitable water vapour changes in the Earth atmosphere toward the target during the observations, in the telluric rest frame shifted for the $V_{sys}$. The magenta horizontal lines show the wavelength range spanned by the planetary motion during the transit for the three lines of the OI triplet. The vertical dotted lines mark the position of the three lines of the triplet.

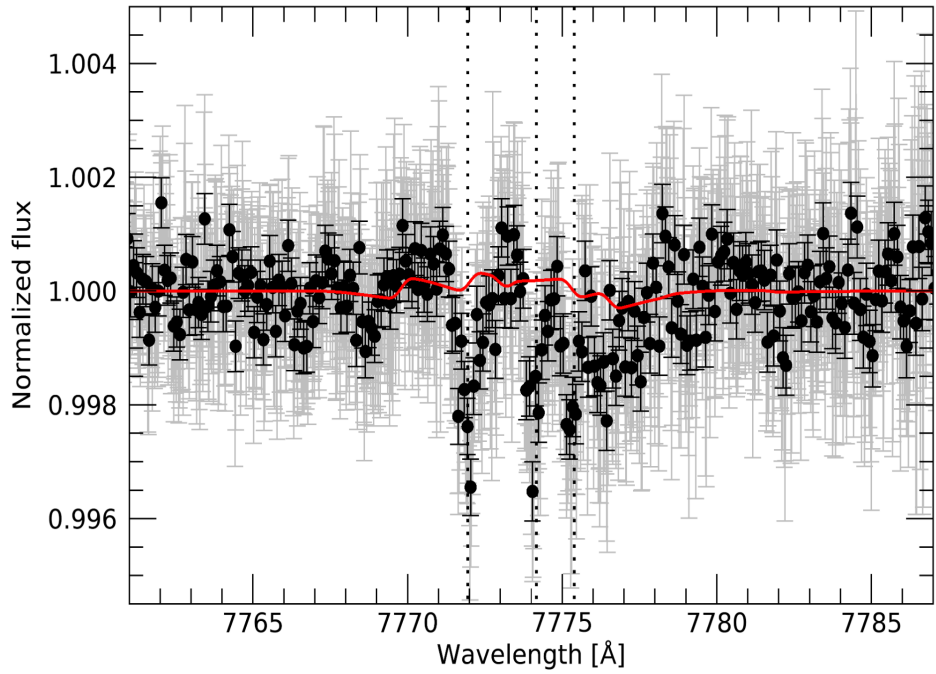

**Supplementary Figure 3.** Raw transmission spectrum of KELT-9b around the OI triplet, before correcting for the stellar contamination of center-to-limb variation and RM effect (red line). The vertical dashed lines indicate the expected position of the OI triplet in the planetary rest frame.

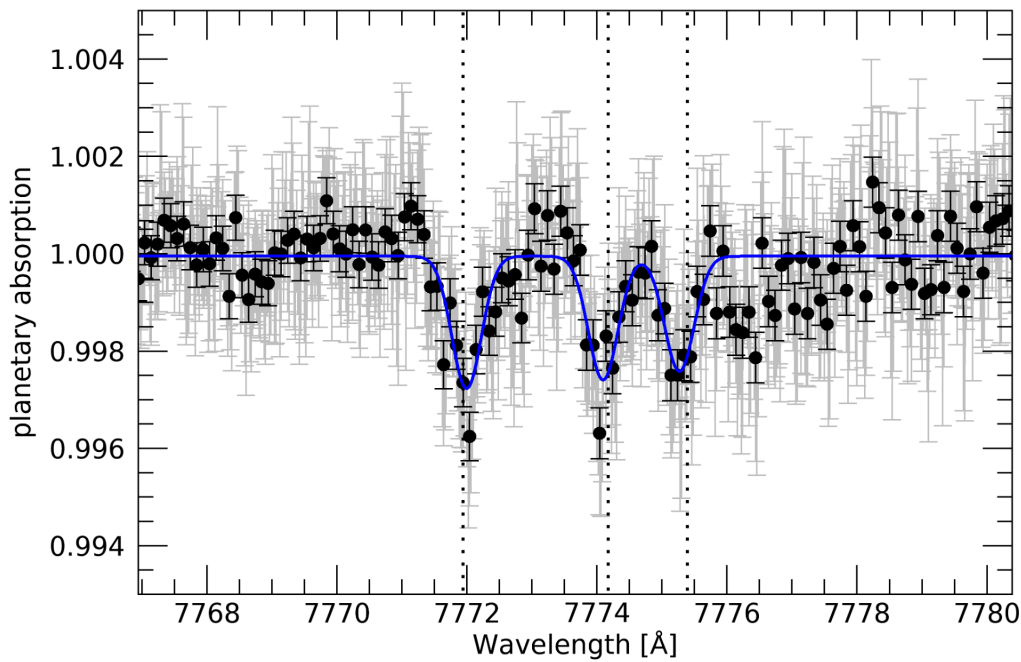

**Supplementary Figure 4.** Transmission spectrum of KELT-9b around the OI triplet. The blue line shows the Gaussian best fits to each line. Black points show an 0.1 Angstrom binning. The vertical dotted lines mark the expected position of the OI triplet lines.

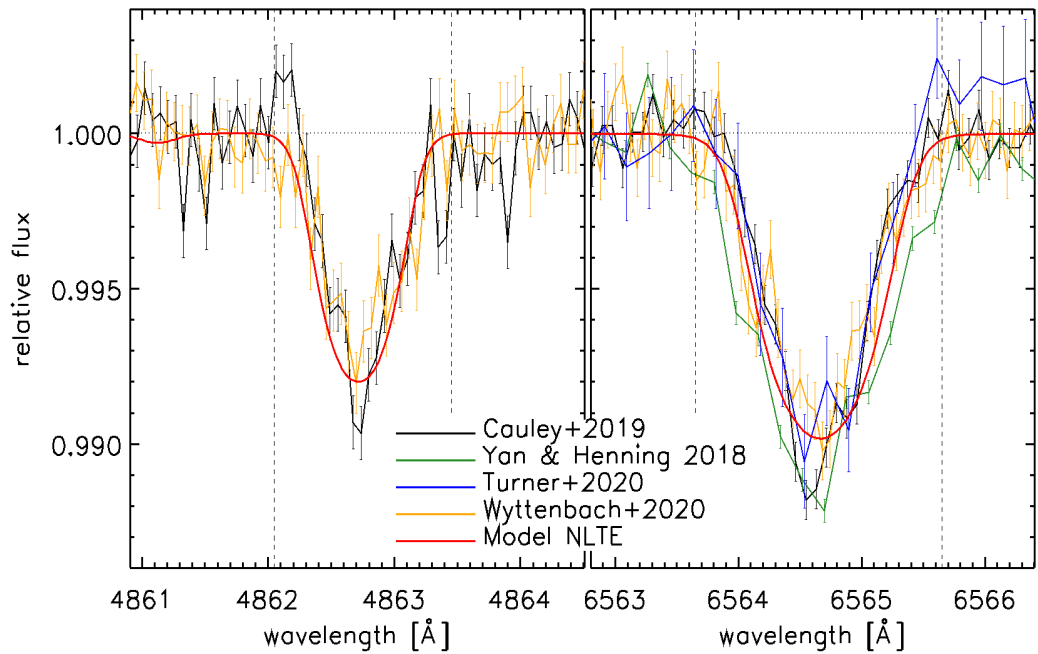

**Supplementary Figure 5.** Comparison between the NLTE synthetic transmission spectrum accounting for $v_{mic}$=3 km s$^{-1}$ and $v_{mac}$=13 km s$^{-1}$ (red) and the observations[9,10,11,12] in the region of the Hα (right) and Hβ (left) Balmer lines.

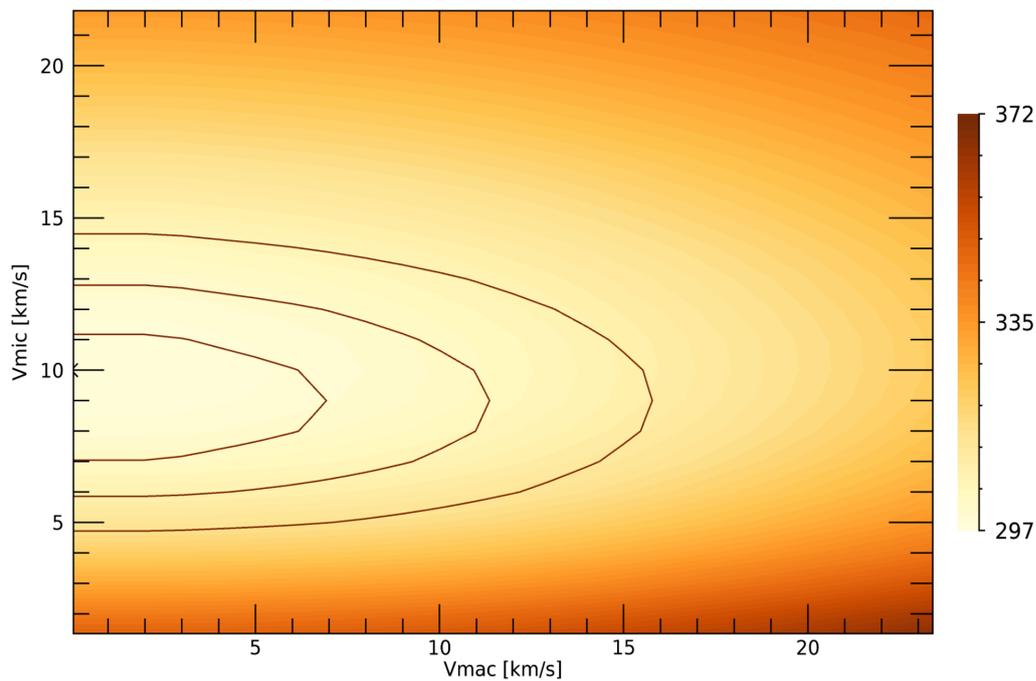

**Supplementary Figure 6.** $X^2$ minimization map of the LTE synthetic spectrum with contour lines of $1\sigma$, $2\sigma$ and $3\sigma$ confidence intervals for the values of the microturbulence ($v_{mic}$) and macroturbulence ($v_{mac}$) velocities.

**Supplementary Table 1.** Summary of the results obtained from the Gaussian fits of the OI triplet. The third line could be partly influenced by water micro-tellurics.

| $\lambda_{air}$ [Angstrom] | contrast [%] | center [km s$^{-1}$] | FWHM [km s$^{-1}$] |
|---|---|---|---|
| 7771.94 | 0.272±0.024 | 2.1±1.1 | 20.7±1.8 |
| 7774.17 | 0.254±0.027 | -2.9±1.1 | |
| 7775.39 | 0.237±0.027 | -4.7±1.2 | |
| triplet | 0.255±0.015 | -1.7±0.7 | 20.7±1.8 |